**Generalized reverberation theory in diffuse sound fields: Introducing mean residual free path for macroscopic and microscopic unification**

Toshiki Hanyu,[1,a]

[1] *Department of Architecture and Living Design, Nihon University Junior College, 7-24-1 Narashinodai, Funabashi, Chiba 274-8501, Japan*

## ABSTRACT

Sabine's foundational theory established the cornerstone of modern architectural acoustics. However, it fails to predict zero reverberation time in perfectly absorptive rooms. Eyring subsequently addressed this issue by proposing a new formula, which Knudsen later extended to incorporate air absorption. Nevertheless, Eyring's underlying approach contains fundamental theoretical contradictions. To resolve these flaws, the author previously introduced macroscopic and microscopic models for a revised reverberation theory; however, their mathematical unification, rigorous derivation from differential equations, and the mechanism behind Eyring's underestimation remained unclarified. This paper presents a comprehensive generalization and mathematical foundation of the revised theory. The macroscopic model is derived directly from fundamental energy differential equations, incorporating air absorption and the mean residual free path. By reformulating the microscopic model as a sequential convolution process, its fundamental consistency with the macroscopic model is mathematically demonstrated. Crucially, it is shown that Eyring's underestimation can be interpreted as stemming universally from the structural omission of temporal variance expansion ($\sigma_n^2 = n\sigma^2$) inherent to multiple reflections, regardless of the assumed probability distribution. Finally, ray-tracing simulations validate the scale-invariant accuracy of the proposed theory. Ultimately, this mathematically consistent framework establishes the theoretical limit for future generalized theories in non-diffuse sound fields.

[a] Email: hanyu.toshiki@nihon-u.ac.jp, toshiki.hanyu@gmail.com

## I. INTRODUCTION

The foundational principles of architectural acoustics were firmly established by W. C. Sabine in the late 1890s [1]. Sabine's reverberation theory is predicated on a diffuse sound field, wherein acoustic energy is uniformly distributed with no directional bias in propagation. This theory provided the crucial physical insight that reverberation time is proportional to room volume and inversely proportional to the equivalent absorption area. However, theoretical issues with this formulation have long been pointed out. Because it was intuitively assumed that reverberation time should be zero in a completely absorptive room (average absorption coefficient $\bar{\alpha} = 1$) due to the absence of reflected sound, the non-zero reverberation time predicted by Sabine's formula was pointed out as a contradiction.

To resolve this contradiction under completely absorptive conditions, C. F. Eyring proposed a new reverberation theory in 1930 that yields a zero reverberation time when $\bar{\alpha} = 1$ [2]. He achieved this by adopting an approach in which the energy decay after wall reflections is expressed as $(1 - \bar{\alpha})^n$, where $\bar{\alpha}$ is the average absorption coefficient and $n$ is the reflection order. Subsequently, V. O. Knudsen incorporated the effect of air absorption [3], and today, the Eyring-Knudsen formula remains widely used as a fundamental framework for room acoustic design. Naturally, in the modern acoustic design of spaces such as concert halls, the direct calculation of reverberation time using Sabine's or Eyring's theory is generally limited to the initial design stages; more accurate sound field predictions typically rely on computer simulations [4–7] or acoustic scale-model experiments [8–12]. However, a robust theoretical foundation is indispensable for the physically correct interpretation of these advanced simulation and experimental results. In this regard, the diffuse field theories of Sabine and Eyring continue to play an essential role as this theoretical bedrock.

Furthermore, since actual indoor spaces are never perfectly diffuse sound fields, various improvements have been attempted to construct reverberation theories for non-diffuse sound fields from multiple perspectives. These include alternative calculation methods for the average absorption coefficient [13], directional reverberation models focusing on rectangular rooms with uneven distributions of absorption [14–21], and the incorporation of diffusing elements and wall scattering [22–25]. Although these advanced efforts remain largely at the research stage, some models have been adopted for specific practical applications, such as the acoustic design of rectangular rooms. Nevertheless, the classical theories of Sabine and Eyring are still widely used today in general acoustic design and standard measurements, such as the reverberation room method for sound absorption [26]. Even if a comprehensively generalized reverberation theory for non-diffuse sound fields is established in the future, this new theory must inherently encompass the reverberation theory for diffuse sound fields as its limit of perfect diffusion. Therefore, establishing a mathematically and physically consistent reverberation theory for the limit state of a diffuse sound field will continue to hold profound significance into the future, serving as a firm guiding principle for the theoretical development of non-diffuse sound fields.

However, this paper reveals that Eyring's theory, which has been regarded as the definitive answer for room acoustics in diffuse sound fields for nearly a century, actually contains fundamental theoretical contradictions. First, introducing Eyring's absorption term into the conventional differential equation based on the room acoustic energy balance [27–30] causes the term to diverge to infinity, leading to the physically impossible conclusion that the boundaries absorb infinite energy. This is not merely a mathematical breakdown but represents a clear failure as a physical model. That is, only Sabine's theory can be derived from the conventional differential equations, making it impossible to deduce Eyring's theory. Second, as will be clarified later in this paper, Eyring's theory lacks strict consistency as a physical theory because the decay rate determining the steady-state energy density does not match the decay rate during the decay process.

In response to these theoretical flaws, the author previously proposed macroscopic [31] and microscopic [32] models based on the new concept of the "reverberation of direct sound" to reconstruct a contradiction-free reverberation theory. However, the initial macroscopic model relied on probabilistic reasoning rather than a rigorous derivation from the fundamental energy continuity equation (differential equation) of the sound field, leaving issues such as the formal incorporation of air absorption unresolved. Furthermore, although the microscopic model showed good numerical agreement with the macroscopic model, the mathematical foundation explaining their structural consistency remained unclarified. Additionally, regarding the mechanism by which Eyring's theory underestimates reverberation time, the previous study merely provided qualitative illustrations using specific distributions (e.g., normal distribution), failing to establish a universal mathematical demonstration applicable to any arbitrary probability distribution.

Therefore, the objective of this paper is to completely establish the mathematical foundation and full scope of the revised reverberation theory through the following three primary breakthroughs:

1. Derivation of a generalized macroscopic model directly from the fundamental energy differential equations, accommodating air absorption and the new spatial parameter.

2. Mathematical demonstration of the structural consistency between the macroscopic model and the microscopic model formulated as a sequential convolution process.

3. Universal mathematical elucidation of the mechanism behind Eyring's underestimation, demonstrating that it stems structurally from the omission of "linear expansion of temporal variance" inherent to sequential multiple reflections, independent of the underlying probability distribution.

The revised reverberation theory presented in this study is not merely aimed at improving engineering calculation accuracy. Its true significance lies in resolving long-standing inconsistencies in the foundational theories of room acoustics and reconstructing a contradiction-free theoretical basis. When a generalized reverberation theory for non-diffuse sound fields—incorporating not only absorption but also diffusion as parameters—is established in the future, the revised theory proposed herein should serve as the theoretical limit for a completely diffuse sound field where diffusion is maximized.

The remainder of this paper is organized as follows. Section II defines the core concept of the reverberation of direct sound and the mean residual free path $\bar{\ell}_d$. Section III explicitly details the mathematical and physical inconsistencies in conventional theories. Sections IV and V derive the generalized revised reverberation theory based on the macroscopic energy density balance and the microscopic sequential convolution of probability distributions, respectively. Section VI conducts a quantitative comparison with conventional theories and demonstrates the structural consistency between the proposed macroscopic and microscopic formulations. Section VII presents a numerical validation using ray-tracing simulations to verify the scale-invariant accuracy of the proposed theory. Finally, Section VIII concludes the study.

## II. CORE CONCEPT: MEAN RESIDUAL FREE PATH

### A. Definition of "Reverberation of Direct Sound"

Reverberation theory in a diffuse sound field is a probabilistic and statistical framework that describes the expected value (spatial average) of the acoustic energy density using the room volume $V$, the room surface area $S$, and the average absorption coefficient $\bar{\alpha}$, without relying on deterministic parameters such as a specific room shape or maximum dimensions. This framework assumes that a perfectly diffuse state of the sound field is maintained during both the steady state and the decay process. Under this premise, this paper first reconsiders the behavior of acoustic energy in a space where the boundaries are completely absorptive ($\bar{\alpha} = 1$).

Historically, based on the intuitive fact that "if no reflected sound is generated, no audible reverberation occurs," the non-zero reverberation time predicted by Sabine's formula at $\bar{\alpha} = 1$ has been regarded as a contradiction. However, returning to the physical definition of reverberation—the decay of the average acoustic energy density in a room—leads to an entirely different interpretation.

Assume that a single omnidirectional sound source exists in a space and stops emitting sound at time $t = 0$ in a steady state. Even under the condition of $\bar{\alpha} = 1$, the acoustic energy in the space does not drop to zero instantaneously. As illustrated in FIG. 1, when the source stops, a silent region of radius $ct$ (where $c$ is the speed of sound) expands spherically from the source. That is, the acoustic energy continues to propagate freely in the space until the last emitted sound energy reaches the boundaries and is completely absorbed.

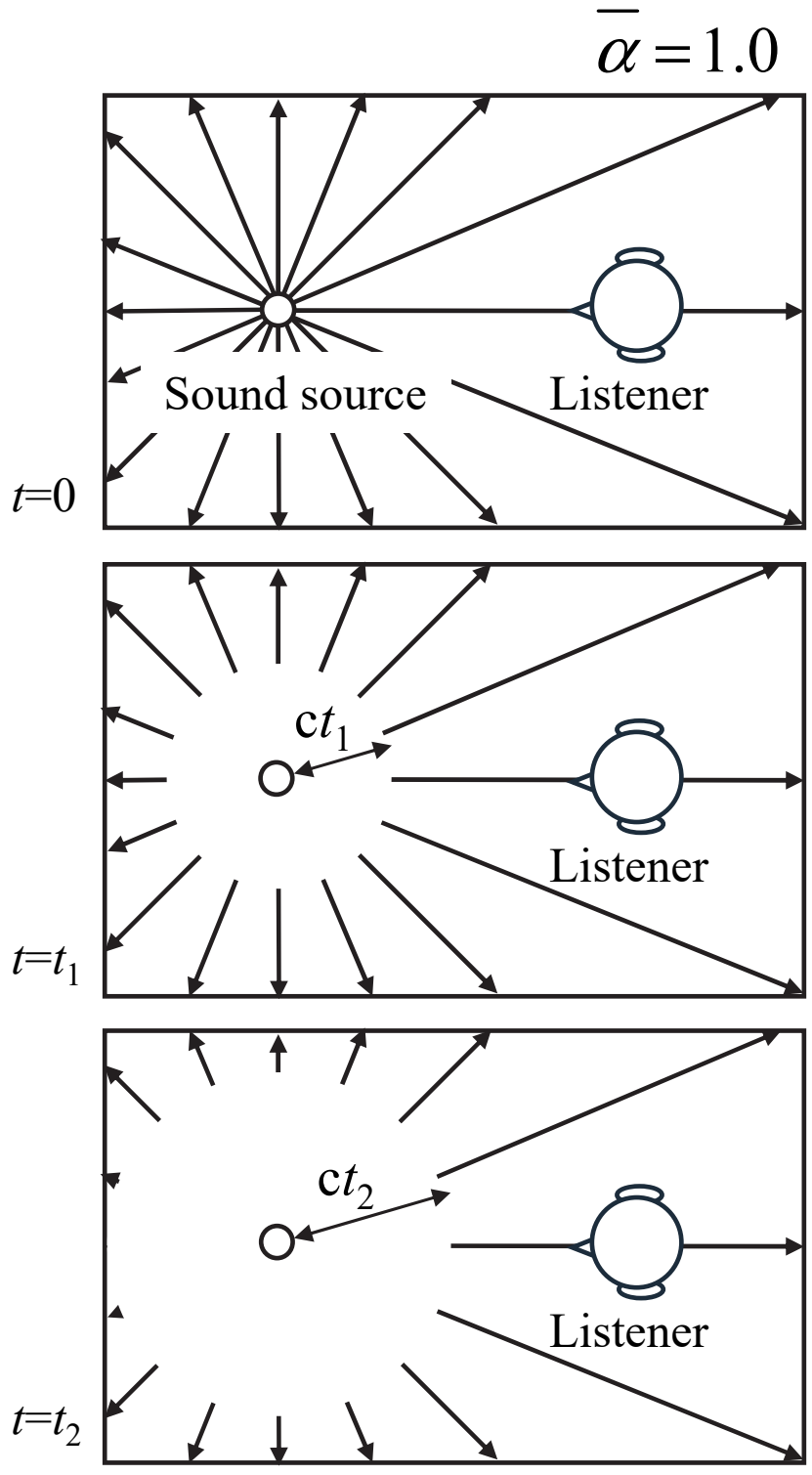

FIG. 1. Conceptual diagram illustrating the propagation and decay of direct sound over time in a room with completely absorptive boundaries ($\bar{\alpha} = 1.0$). (Adapted with permission from Ref. 33. Copyright 2025 Acoustical Society of Japan.)

This phenomenon is also derived from a probabilistic and statistical interpretation in which a diffuse sound field is simulated by an infinite number of sound sources (FIG. 2). When these infinite sources stop emitting sound simultaneously in a steady state, a listener in this idealized field would not perceive all sounds ceasing at once. Instead, due to the varying arrival times proportional to the distances from each source to the listener, the sound would be perceived as fading out gradually. In other words, in a theoretical diffuse sound field, even under the condition of $\bar{\alpha} = 1$, the decay of sound mathematically occurs and should be audibly perceived. In a real room with a single source as illustrated in FIG. 1, this decay cannot be perceived as reverberation; the failure to distinguish between everyday auditory experience and the theoretical phenomenon of a diffuse field caused this purely physical reality to be overlooked, hindering the fundamental revision of reverberation theory for nearly a century. Therefore, this shift in perspective is an essential first step in unraveling long-standing theoretical contradictions and reconstructing reverberation theory based on physical reality.

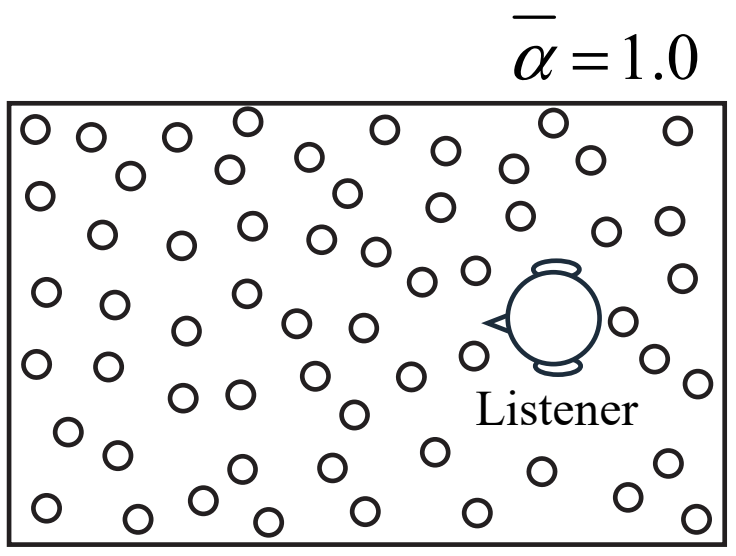

FIG. 2. Conceptual illustration of a diffuse sound field, represented as an aggregate of direct sounds from multiple sound sources under completely absorptive conditions ($\bar{\alpha} = 1.0$). (Adapted with permission from Ref. 33. Copyright 2025 Acoustical Society of Japan.)

### B. Extension to Reflected Sound Fields: Mean Residual Free Propagation Time

The "reverberation of direct sound" defined in the previous section is not limited to the special case of $\bar{\alpha} = 1$ but is a concept that should be universally applied to reflected sound fields with an arbitrary average absorption coefficient. During a reverberation process involving reflections, the acoustic energy reflected at a boundary propagates freely through the space again at a constant speed $c$ until it reaches the next boundary.

As illustrated in FIG. 3, focusing on an arbitrary "sound particle" (or an infinitesimal wave packet of acoustic energy) propagating through the space, the time it takes to reach the next boundary from its current, arbitrary position is defined as the residual free propagation time. In other words, whether it is direct sound or reflected sound, acoustic energy in a "propagating state" undergoes an essentially equivalent process of free propagation until it reaches a boundary to be absorbed or re-reflected. Therefore, to accurately describe the decay process of the total reverberation energy in a diffuse sound field, it is strictly necessary to consider not only the energy loss due to boundary absorption but also the probabilistic and statistical distribution of this "mean residual free propagation time" within the space. Furthermore, as visually depicted in FIG. 3, individual sound particles travel specific distances—such as $\ell_{d1}$ to $\ell_{d6}$—during these residual free propagation times before striking a boundary. The statistical average of these infinite individual paths constitutes the "mean residual free path" $\bar{\ell}_d$, a new spatial parameter that will be formally introduced and mathematically formulated in Sec. II C.

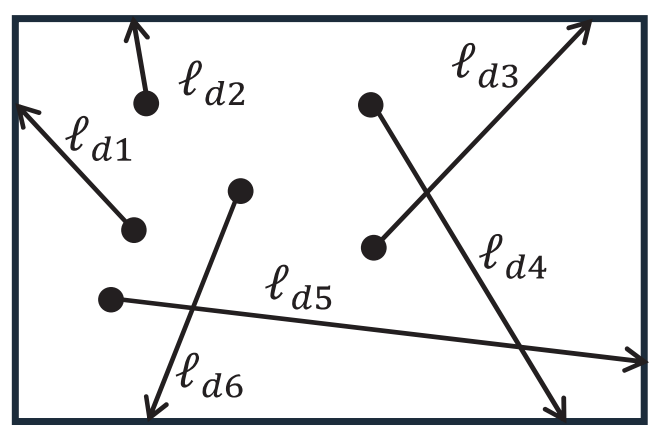

FIG. 3. Conceptual illustration of the residual free paths, which are used to determine the mean residual free propagation time. The paths $\ell_{d1}$ to $\ell_{d6}$ represent the distances traveled by individual

sound particles before their first boundary collision. The ensemble average of these infinite paths, defined as the mean residual free path $\bar{\ell}_d$, is detailed in Sec. II C. (Adapted with permission from Ref. 33. Copyright 2025 Acoustical Society of Japan.)

### C. Introduction of New Parameter $\bar{\ell}_d$: Mean Residual Free Path

To incorporate the residual free propagation time of acoustic energy into mathematical models, the parameter $\bar{\ell}_d$ is introduced in this study.

The conventional mean free path, $\bar{\ell} = 4V/S$ [34, 35], which is widely used in room acoustics, represents the expected distance that sound departing from an arbitrary point on a boundary travels before reaching the next boundary (i.e., the wall-to-wall distance). In contrast, $\bar{\ell}_d$ defined in this study represents the expected distance that direct sound emitted from an arbitrary point in the room (the sound source position) travels before reaching a boundary. Since a sound source is generally located inside the room rather than on a boundary, the relationship $\bar{\ell}_d < \bar{\ell}$ holds from a probabilistic and statistical perspective.

More importantly, $\bar{\ell}_d$ is not a parameter exclusively applicable to direct sound. Recalling the individual propagation paths ($\ell_{d1}$ to $\ell_{d6}$) of sound particles previously illustrated in FIG. 3, during the steady state or the reverberation decay process, the expected value of the "remaining distance" that any propagating acoustic energy (including reflected sound) will travel from its current position to the next boundary is also equivalent to $\bar{\ell}_d$. Accordingly, the mean residual free propagation time discussed in the previous section is explicitly formulated as $\bar{\ell}_d/c$, where $c$ is the speed of sound. Consequently, $\bar{\ell}_d$ functions as the "mean residual free path," uniformly describing the spatial propagation characteristics of acoustic energy in a diffuse sound field from both macroscopic and microscopic perspectives, independent of specific room shapes. The mathematical formulations developed in the subsequent sections are formulated using this new parameter $\bar{\ell}_d$.

## III. THEORETICAL INCONSISTENCIES AND MATHEMATICAL FLAWS IN CONVENTIONAL FORMULATIONS

### A. Inconsistency Between Steady-State Energy Density and Decay Rate in Eyring Theory

An absolute requirement for constructing a physically consistent reverberation theory is that the time decay rate determining the steady-state acoustic energy density and the time decay rate governing the reverberation process after the sound source stops must be defined as the identical parameter. This requirement is deductively derived from the following mathematical process.

Consider a situation in a room of volume $V$ where a sound source with an acoustic power $W$ emits sound for a brief period $\Delta t$. Assuming the decay rate per unit time (time decay rate) of the acoustic energy density is $\lambda$, the decay of the acoustic energy density is expressed as:

$$E(t) = \frac{W\Delta t}{V}\exp(-\lambda t) \qquad (1)$$

By taking the limit as $\Delta t \to dt$, the average energy impulse response in the room is obtained as:

$$dE(t) = \frac{W}{V}\exp(-\lambda t)\, dt \qquad (2)$$

Applying Schröder integration [36] to this impulse response yields the reverberation decay curve after the source stops from a steady state:

$$E(t) = \int_t^{\infty} \frac{W}{V}\exp(-\lambda \tau)\, d\tau = \frac{W}{V\lambda}\exp(-\lambda t) \qquad (3)$$

By substituting $t = 0$ into Eq. (3), the average acoustic energy density in the steady state, $E_0$, is determined as $E_0 = W/(V\lambda)$. That is, both the steady-state energy density $E_0$ and the decay process $\exp(-\lambda t)$ must be mathematically governed by the identical $\lambda$.

In Sabine's theory, the time decay rate is defined as $\lambda = cS\bar{\alpha}/(4V)$ (where $c$ is the speed of sound, $S$ is the surface area, and $\bar{\alpha}$ is the average absorption coefficient). Applying this to Eq. (3) demonstrates that the identical decay rate is used for both the steady state and the decay process, fully preserving theoretical and mathematical consistency.

However, a profound theoretical inconsistency exists in Eyring's theory. The decay rate for the reverberation process in Eyring's theory is derived as $\lambda = -cS\ln(1-\bar{\alpha})/(4V)$. Nevertheless, for the steady-state energy density $E_0$, the exact same $\lambda = cS\bar{\alpha}/(4V)$ as in Sabine's theory is assumed [2]. In other words, Eyring's formulation employs two different decay rates to establish the steady state and to describe the decay process within a framework describing the identical physical space. This fact mathematically indicates that Eyring's theory, which is widely used today, lacks strict consistency as a physical theory.

### B. Breakdown of Boundary-Incidence Energy Balance Equations

A differential equation based on the acoustic energy balance in a room is utilized as the physical foundation of reverberation theory [27–30]. As illustrated in FIG. 4, let $I_w$ be the acoustic energy incident on the boundary per unit time and unit area (incident sound intensity). The time derivative of the acoustic energy in the room, $V(dE/dt)$, is described by the following equation:

$$V\frac{dE}{dt} = W - I_w S\bar{\alpha} \qquad (4)$$

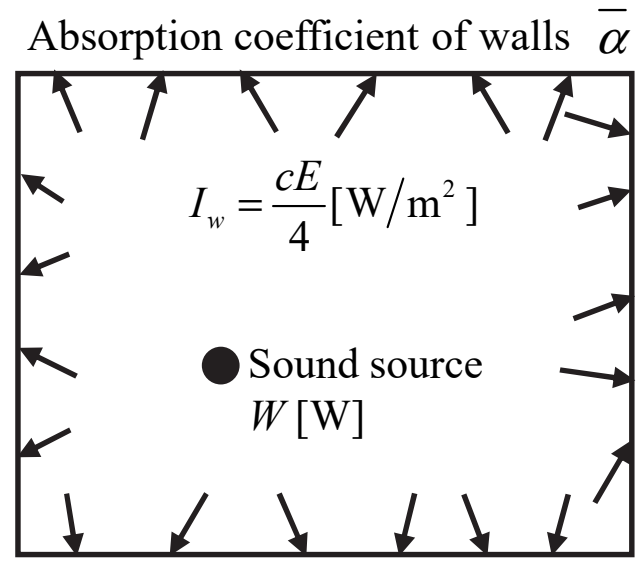

FIG. 4. Conceptual diagram illustrating the incident sound intensity on boundaries, which forms the basis of the conventional differential equation. (Reproduced with permission from Ref. 33. Copyright 2025 Acoustical Society of Japan.)

Substituting the relationship in a diffuse sound field, $I_w = cE/4$, into Eq. (4) yields the standard differential equation that forms the basis of conventional reverberation theories:

$$V\frac{dE}{dt} = W - \frac{cE}{4}S\bar{\alpha} \qquad (5)$$

In the steady state ($dE/dt = 0$), Eq. (5) yields the steady-state energy density $E_0 = 4W/(cS\bar{\alpha})$. Furthermore, solving Eq. (5) under the condition that the sound source stops ($W = 0$) from the steady state naturally derives Sabine's decay curve. That is, the conventional differential equation possesses a structure focusing on the "incident energy" to the boundaries as illustrated in FIG. 4, and from this, only Sabine's theory can be deductively derived.

### C. Divergence Issue of Eyring's Absorption Term in Differential Equations

If Eyring's theory is a physically valid model, it must also be deductively derived from a differential equation based on energy balance. Introducing Eyring's absorption term, $-\ln(1-\bar{\alpha})$, instead of the absorption coefficient $\bar{\alpha}$ based on Eyring's theory results in the following differential equation:

$$V\frac{dE}{dt} = W - \frac{cE}{4}S[-\ln(1-\bar{\alpha})] \qquad (6)$$

When the completely absorptive condition ($\bar{\alpha} \to 1$) is applied here, Eyring's absorption term diverges to $-\ln(1-\bar{\alpha}) \to \infty$; therefore, the second term on the right side of Eq. (6) (the energy absorbed at the boundaries) diverges to infinity. This implies a physically impossible conclusion: "in a space with finite energy, the boundaries absorb infinite energy". This result indicates that Eq. (6) clearly breaks down as a physical model, exceeding the limits of a mere mathematical approximation. Therefore, it is impossible to deductively derive Eyring's theory from the differential equation.

### D. Incompatibility with Air Absorption Parameter

Furthermore, another structural flaw exists in the conventional differential equations, such as Eqs. (4) and (5). It is the problem that the effect of air absorption (spatial propagation loss) cannot be intrinsically incorporated.

The air absorption parameter introduced by Knudsen [3] represents the energy loss proportional to the distance propagated through space. However, as illustrated in FIG. 4, because the differential equation in Eq. (4) is constructed relying solely on the "incident energy $I_w$ at the boundaries," the attenuation during propagation within the space cannot be naturally incorporated into its mathematical structure. The Eyring-Knudsen formula, which is widely used today, is formed by adding the air absorption term *a posteriori* to the boundary absorption term, rather than being uniformly derived from a differential equation. This incompatibility strongly suggests that the physical model of reverberation theory should be fundamentally reconstructed based on the "propagating sound intensity" through space, rather than the "incidence on boundaries".

## IV. GENERALIZED MACRO MODEL: PROPAGATING INTENSITY-BASED DERIVATION

### A. New Differential Equation Based on Propagating Sound Intensity

To overcome the structural limitations of conventional theories pointed out in Chapter III, this section constructs a new differential equation focusing on the balance of the "propagating sound intensity" through space, rather than the "incident energy on boundaries".

As a premise, it is assumed that a perfectly diffuse sound field is established in a room of volume $V$. Let $W$ be the total acoustic energy supplied to this system per unit time, and $\lambda$ be the time decay rate of the acoustic energy density. The decrease in energy density per unit time is expressed as $\lambda E$. Therefore, the time derivative of the total acoustic energy, $V(dE/dt)$, in the space is formulated as follows:

$$V\frac{dE}{dt} = W - \lambda VE \qquad (7)$$

Here, the time decay rate $\lambda$ is converted into the decay rate with respect to the spatial propagation distance (spatial decay rate) $\delta$, using the relationship $\lambda = c\delta$. Furthermore, by applying the relationship between the sound intensity $I$ propagating through space and the energy density $E$ in a diffuse sound field ($I = cE$), Eq. (7) is rewritten as:

$$V\frac{dE}{dt} = W - \delta VI \qquad (8)$$

Equation (8) is a new differential equation that breaks away from the dependence on the boundary incident intensity $I_w$ in the conventional Eq. (4) and describes the energy balance using the sound intensity $I$ propagating within the space as a direct variable. As conceptualized in FIG. 5, this shift from boundary incidence to spatial propagation forms the fundamental physical framework of the revised macro model.

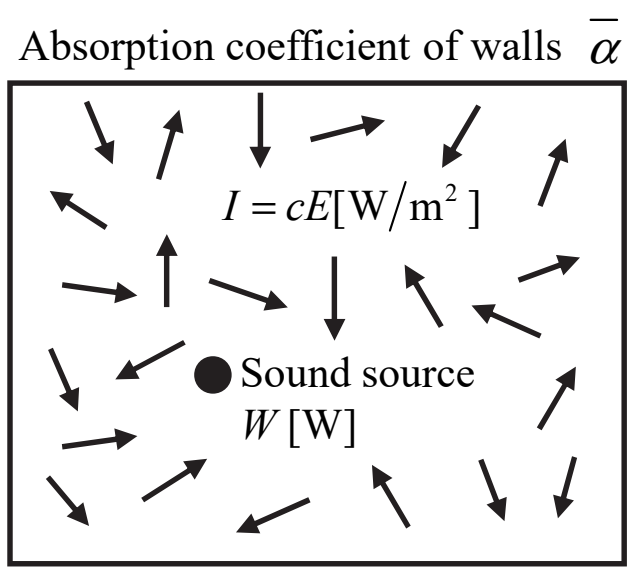

FIG. 5. Conceptual illustration of the revised differential equation based on the propagating sound intensity within the room. (Reproduced with permission from Ref. 33. Copyright 2025 Acoustical Society of Japan.)

### B. Mean Absorption Free Path $\bar{\ell}_a$ and Integration of Air Absorption

The spatial decay rate $\delta$ in Eq. (8) consists of two factors: air absorption during propagation and boundary absorption. Let $m$ be the air absorption rate per unit propagation distance. To evaluate the spatial decay rate due to boundary absorption, a new parameter, the "mean absorption free path" $\bar{\ell}_a$, is introduced.

To strictly define $\bar{\ell}_a$, this study assumes a "sound particle" as the indivisible minimum unit of acoustic energy. If acoustic energy were treated as a continuous quantity, it would undergo a continuous decay process where the energy is multiplied by $(1-\bar{\alpha})$ at each boundary reflection. Since this process has no endpoint where the energy is "completely absorbed," the concept of the "distance traveled until absorption" cannot be mathematically established. In contrast, assuming an indivisible sound particle, it follows a probabilistic process upon collision with a boundary: it is either completely absorbed (destroyed) with a probability of the average absorption coefficient $\bar{\alpha}$, or reflected retaining its energy with a probability of $(1-\bar{\alpha})$.

Based on this premise, $\bar{\ell}_a$ is defined as the expected value of the total distance a sound particle travels without being absorbed, from the time it is emitted into the space until it is probabilistically absorbed and destroyed. Its reciprocal, $1/\bar{\ell}_a$, represents the decay rate per unit propagation distance due to boundary absorption. Therefore, $\delta$ is given by:

$$\delta = m + \frac{1}{\bar{\ell}_a} \qquad (9)$$

The mean absorption free path $\bar{\ell}_a$ is mathematically derived. An arbitrary sound particle travels the mean residual free path $\bar{\ell}_d$ before reaching the first boundary. Upon collision with the boundary, the probability of being reflected without absorption is $(1 - \bar{\alpha})$, and thereafter, it repeats reflections at intervals of the conventional mean free path $\bar{\ell} = 4V/S$. Therefore, by taking the expected value (infinite series sum) of the propagation distance according to the number of reflections $n$, $\bar{\ell}_a$ is formulated as follows:

$$\bar{\ell}_a = \bar{\ell}_d + \bar{\ell} \sum_{n=1}^{\infty} (1 - \bar{\alpha})^n = \bar{\ell}_d + \bar{\ell} \frac{1 - \bar{\alpha}}{\bar{\alpha}} \qquad (10)$$

By substituting Eqs. (9) and (10) into Eq. (8), a generalized differential equation is completed in which air absorption and boundary absorption are perfectly integrated as a mathematical structure.

### C. Derivation of Revised Reverberation Time Formula Incorporating Mean Residual Free Path

For the mean absorption free path $\bar{\ell}_a$ derived in Eq. (10), the conventional mean free path widely known in room acoustics, $\bar{\ell} = 4V/S$, is substituted. Furthermore, introducing the room constant $R = S\bar{\alpha}/(1 - \bar{\alpha})$ as a parameter governing the effect of reflected sound simplifies the second term of Eq. (10) (the contribution of the reflected sound propagation distance) as follows:

$$\bar{\ell} \frac{1 - \bar{\alpha}}{\bar{\alpha}} = \frac{4V}{S} \frac{1 - \bar{\alpha}}{\bar{\alpha}} = \frac{4V}{R} \qquad (11)$$

Therefore, the mean absorption free path $\bar{\ell}_a$ is expressed in a highly transparent form using the unknown mean residual free path $\bar{\ell}_d$ and the room constant $R$:

$$\bar{\ell}_a = \bar{\ell}_d + \frac{4V}{R} \qquad (12)$$

By substituting this into Eq. (9), the time decay rate $\lambda$ in the revised theory is deductively derived as:

$$\lambda = c\left(m + \left(\bar{\ell}_d + \frac{4V}{R}\right)^{-1}\right) \qquad (13)$$

This time decay rate $\lambda$ consistently describes both the steady-state energy density and the reverberation decay process. The reverberation time $T$ is defined as the time required for the acoustic energy density to decay by 60 dB (i.e., to $10^{-6}$ of its initial value). By solving the exponential decay equation $\exp(-\lambda T) = 10^{-6}$, the reverberation time is uniquely derived as $T = 6 \ln 10 / \lambda \approx 13.8/\lambda$. Substituting Eq. (13) into this relationship yields the new revised reverberation time formula, $T_{\mathrm{Revised}}$:

$$T_{\text{Revised}} = \frac{13.8}{c\left(m + \left(\bar{\ell}_d + \frac{4V}{R}\right)^{-1}\right)} \qquad (14)$$

This Eq. (14) represents the completed form of the macro model that strictly and uniformly describes the reverberation time in a diffuse sound field by incorporating the mean residual free path $\bar{\ell}_d$ and the room constant $R$, without assuming a specific room shape.

### D. Separation of Direct and Reflected Energy: Resolution of Sabine's Paradox

The most significant physical achievement of this formulation is that the decay process of the total reverberation energy can be strictly separated into direct sound and reflected sound. Assuming air absorption is ignored ($m = 0$) for simplicity, the reverberation decay of the total energy $E(t)$ after the sound source stops from a steady state is expressed as:

$$E(t) = \frac{W}{Vc}\left(\bar{\ell}_d + \frac{4V}{R}\right)\exp\left[-ct\left(\bar{\ell}_d + \frac{4V}{R}\right)^{-1}\right] \qquad (15)$$

Applying the completely absorptive condition ($\bar{\alpha} = 1$, i.e., $R \to \infty$) to Eq. (15) extracts the reverberation decay of only the direct sound, $E_d(t)$, where no reflected sound exists:

$$E_d(t) = \frac{W}{V}\frac{\bar{\ell}_d}{c}\exp\left(-\frac{ct}{\bar{\ell}_d}\right) \qquad (16)$$

As shown in Eq. (16), the decay of direct sound is governed not by the total reverberation decay rate $\lambda$, but by its own unique decay rate $c/\bar{\ell}_d$ dependent on the mean residual free path $\bar{\ell}_d$. By subtracting the direct sound decay $E_d(t)$ from the total reverberation energy $E(t)$, the reverberation decay of only the reflected sound, $E_r(t)$, is derived as follows:

$$E_r(t) = \frac{W}{Vc}\left[\left(\bar{\ell}_d + \frac{4V}{R}\right)\exp\left(-ct\left(\bar{\ell}_d + \frac{4V}{R}\right)^{-1}\right) - \bar{\ell}_d\exp\left(-\frac{ct}{\bar{\ell}_d}\right)\right] \qquad (17)$$

Substituting $t = 0$ into Eq. (17) yields the steady-state reflected sound energy density $E_{r0} = \frac{W}{Vc}\left(\frac{4V}{R}\right) = \frac{4W}{cR}$, perfectly satisfying the known theoretical requirements of room acoustics. Furthermore, it is noteworthy that substituting $\bar{\alpha} = 1$ ($R \to \infty$) into Eq. (17) mathematically demonstrates that $E_r(t) = 0$, meaning no reflected sound is generated whatsoever.

Historically, the behavior of Sabine's theory, where the reverberation time does not become zero at $\bar{\alpha} = 1$, has been regarded as a theoretical flaw (paradox). Sabine's formula itself is incomplete as a macro model in that it describes direct and reflected sound using an identical decay rate without separating them. However, the result of the reverberation time at $\bar{\alpha} = 1$ suggested by the formula was not a mere mathematical contradiction, but it incompletely captured the pure physical reality shown in Eq. (16): the "spatial retention of direct sound (free propagation caused by the mean residual free path $\bar{\ell}_d$)". Through this new formulation that explicitly separates direct and reflected sound, Sabine's paradox, which has plagued reverberation theory for a century, is mathematically and physically resolved completely.

## V. GENERALIZED MICRO MODEL: SEQUENTIAL CONVOLUTION APPROACH

While macroscopic reverberation theories based on differential equations of acoustic energy balance provide a broad perspective, capturing the underlying mechanics of the reverberation process requires a microscopic view focusing on individual reflection events. Polack demonstrated this by applying the theory of ergodic dynamical systems (billiards) to room acoustics, concluding that the reverberation process is governed by the probability distribution of the number of reflections a sound ray undergoes over time [37]. He showed that the energy decay function can be formulated directly by summing the probabilities of reflection numbers. However, because his approach primarily aimed to generalize Sabine's theory through a probabilistic interpretation assuming instant mixing, it did not address the physical contradictions inherent in Sabine's and Eyring's theories under highly absorbent conditions.

Independently, Hanyu et al. proposed a stochastic approach to capture the transient response of diffuse sound fields by focusing on the sequential reflection process [38]. In that early work, the time distribution of each reflection order was expressed through the sequential convolution of normal distributions. Although this approach introduced the concept of convolution across reflection orders, the fundamental flaw in Eyring's theory—specifically, the lack of temporal variance expansion—was not recognized by the author at that time. Consequently, that formulation remained a conceptual framework and was not fully developed into a revised reverberation theory.

Building upon these previous studies, this chapter reconstructs a microscopic reverberation model based on reflection orders. By incorporating the concept of the "reverberation of direct sound" alongside temporal variance expansion, a mathematically and physically consistent formulation is established.

### A. Sequential Convolution Formulation and Reverberation Decay Model

In contrast to the macroscopic approach presented in Chapter IV, this chapter reconstructs the reverberation decay from the bottom up by treating the propagation of acoustic energy (sound particles) between boundaries as a stochastic process, summing the temporal distributions at each reflection order.

Let $p_d(t)$ be the probability density function (PDF) of the propagation time for the direct sound emitted from the source to reach the first boundary, and let its mean be the mean residual free propagation time $\bar{\ell}_d/c$. Subsequently, let $p(t)$ be the PDF of the free propagation time per reflection as the sound particle repeatedly reflects between boundaries, and let its mean be the conventional mean free propagation time $\overline{\ell}/c$. Based on the theorem of the sum of independent random variables, the probability density function $P_n(t)$ of the total propagation time for a sound particle experiencing $n$ boundary reflections is rigorously formulated as the sequential convolution integral of the individual PDFs:

$$P_n(t) = p_d(t) * p(t) * \ldots * p(t) \quad (n \text{ times}) \qquad (18)$$

where the asterisk (*) denotes the convolution integral operation. $P_n(t)$ in Eq. (18) is a probability density function; thus, its time integral is unity. Because the mean free propagation time per reflection is $\overline{\ell}/c$, summing $P_n(t)$ over all reflection orders converges to a constant value of $c/\overline{\ell}$ in the steady state ($t \gg 0$), which corresponds to the number of reflections (arrival frequency) per unit time, as illustrated in FIG. 6.

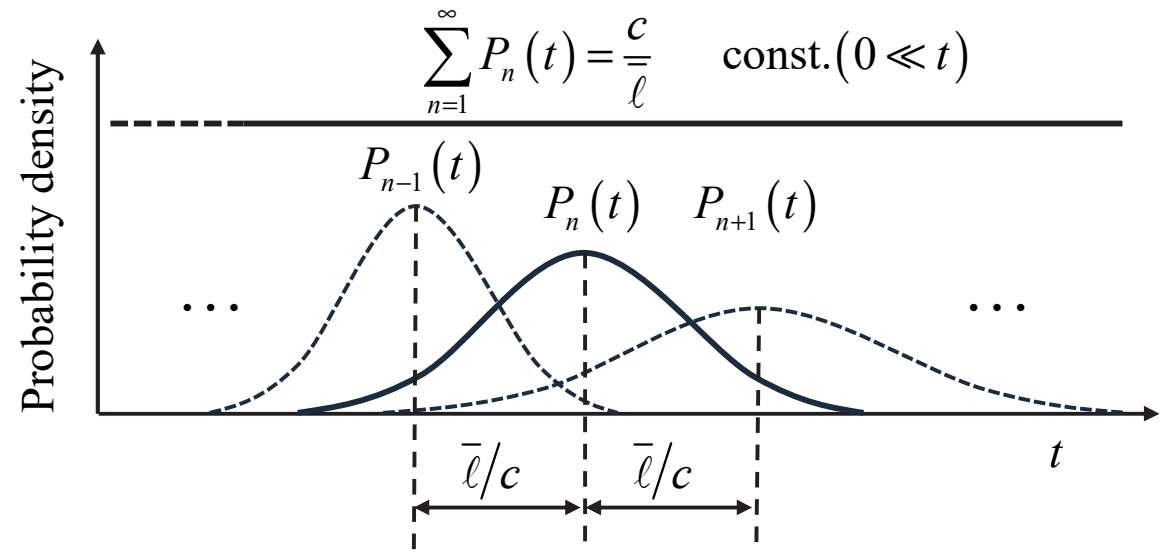

FIG. 6. Conceptual diagram of the probability density of the time distribution of the nth-order reflected sound.

Therefore, to correctly evaluate the energy of the entire system, it is necessary to use $(\overline{\ell}/c)P_n(t)$, which is $P_n(t)$ multiplied by the mean free propagation time $\overline{\ell}/c$, so that the sum over all reflection orders equals unity (1) in the steady state ($t \gg 0$), as shown in FIG. 7. This expresses the exact energy proportion of the $n$-th order reflected sound at time $t$.

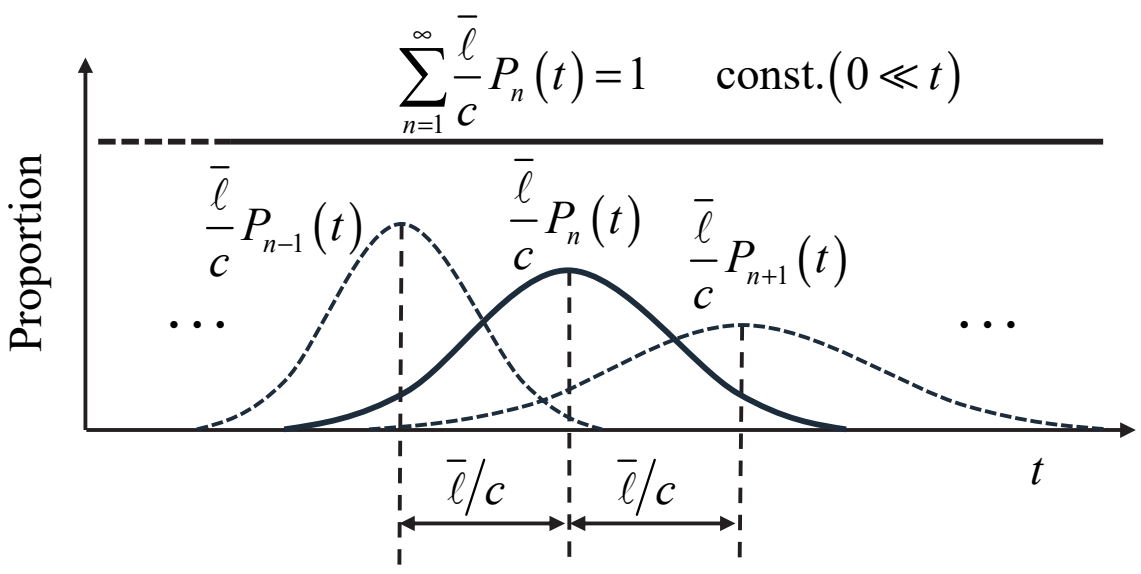

FIG. 7. Conceptual diagram of the time variation of the proportion of the nth-order reflected sound.

Assuming that acoustic energy is emitted from a source with sound power $W$ in a room of volume $V$ during a microscopic time $\Delta t$, it propagates through the space while decaying by a factor of $(1 - \overline{\alpha})$ at each boundary reflection. The temporal energy distribution of the $n$-th order reflected sound, $\Delta E_n(t)$, is formulated by multiplying the energy proportion $(\bar{\ell}/c)P_n(t)$ by the total emitted energy $W\Delta t/V$ and the energy decay rate $(1-\bar{\alpha})^n$:

$$\Delta E_n(t) = \frac{W\Delta t}{V}(1-\bar{\alpha})^n \frac{\bar{\ell}}{c} P_n(t) \qquad (19)$$

By setting $\Delta t \to 0$, the continuous energy impulse response (the rate of change of energy density) for the $n$-th order reflection is derived as:

$$\frac{dE_n(t)}{dt} = \frac{W}{V}(1-\bar{\alpha})^n \frac{\bar{\ell}}{c} P_n(t) \qquad (20)$$

Similarly, the energy impulse response for the direct sound ($n = 0$) is explicitly given by:

$$\frac{dE_d(t)}{dt} = \frac{W}{V}\frac{\bar{\ell}_d}{c} p_d(t) \qquad (21)$$

By applying the Schroeder integration shown in FIG. 8 to this energy impulse response, the reverberation decay $E_n(t)$ of the $n$-th order reflected sound alone from the steady state is precisely given by the following equation:

$$E_n(t) = \int_t^\infty \frac{dE_n(\tau)}{d\tau} d\tau = \frac{W}{V}\frac{\bar{\ell}}{c}(1-\bar{\alpha})^n \int_t^\infty P_n(\tau) d\tau \qquad (22)$$

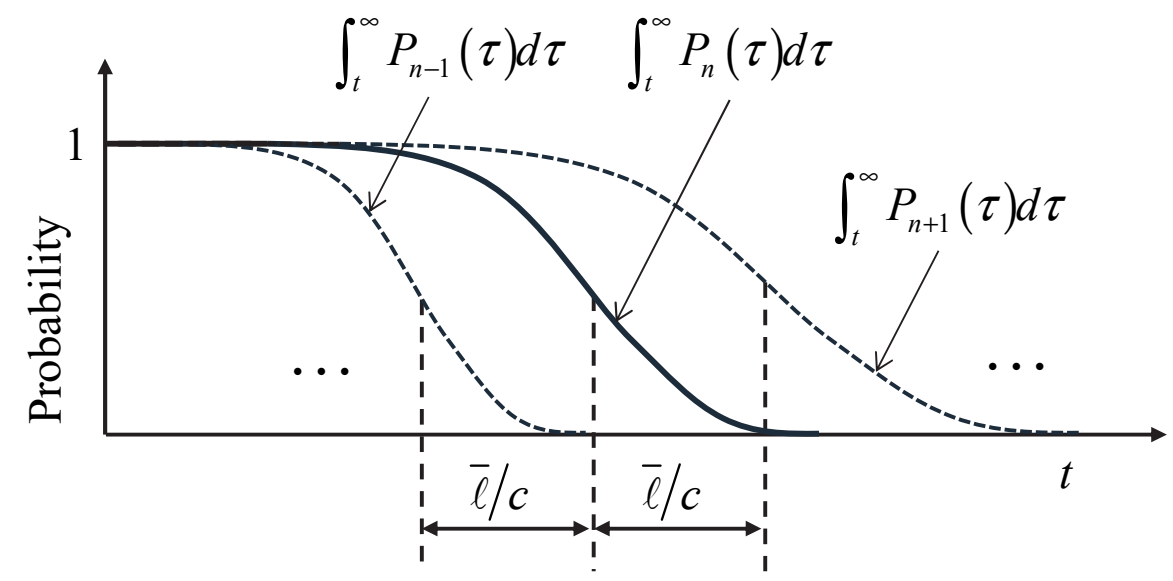

FIG. 8. Schröder integration of the probability density of the nth-order reflected sound.

Finally, to obtain the overall reverberation decay curve $E(t)$ of the entire sound field, it is mathematically essential to sum not only all the reflection orders of the reflected sound but also the decay of the direct sound $E_d(t)$. Therefore, the generalized micro model is constructed as follows:

$$E(t) = E_d(t) + \sum_{n=1}^{\infty} E_n(t) = \frac{W}{V}\frac{\bar{\ell}_d}{c}\int_t^\infty p_d(\tau)d\tau + \frac{W}{V}\frac{\bar{\ell}}{c}\sum_{n=1}^{\infty}(1-\bar{\alpha})^n \int_t^\infty P_n(\tau)d\tau \qquad (23)$$

Equation (23) represents a generalized micro model that directly derives macroscopic reverberation decay from the microscopic stochastic processes of sound particle propagation and absorption, without *a priori* assuming a specific decay curve (such as an exponential function).

While Eq. (23) provides a generalized framework independent of specific probability distributions for the reflected sound, the exact probability density function for the direct sound, $p_d(t)$, can be uniquely identified to satisfy the macroscopic requirement. By equating the direct sound decay term in Eq. (23) with the macroscopic derivation in Eq. (16), the relationship $\int_t^\infty p_d(\tau)d\tau = \exp(-ct/\bar{\ell}_d)$ is established. By differentiating both sides with respect to time $t$, $p_d(t)$ is explicitly mathematically derived as an exponential distribution:

$$p_d(t) = \frac{c}{\bar{\ell}_d} \exp\left(-\frac{c}{\bar{\ell}_d} t\right) \qquad (24)$$

### B. Physical Origin of Temporal Variance Expansion and Limitation of Eyring's Theory

The sequential convolution formulated in Eq. (18) mathematically describes the shape variation of the arrival time probability distribution as the number of reflections $n$ increases. As depicted in the conceptual diagram of FIG. 9, the mean of the distribution shifts by $\bar{\ell}/c$ for each reflection, and the mean arrival time of the total propagation time becomes $\mu_n = \bar{\ell}_d/c + n\bar{\ell}/c$. Here, assuming a diffuse sound field where the propagation times between reflections are mutually independent, the additivity of variance for independent random variables holds regardless of the specific shape of the base distribution. Consequently, the variance of the total arrival time, $\sigma_n^2$, is strictly given by the following equation:

$$\sigma_n^2 = \sigma_d^2 + n\sigma^2 \qquad (25)$$

Mean of direct sound distribution
Mean of 1st-order reflected sound distribution
$t$=0
$\bar{\ell}_d/c$
$\bar{\ell}/c$
$\bar{\ell}/c$
$t$

FIG. 9. Conceptual diagram of the mean arrival times for the direct and nth-order reflected sounds.

The crucial physical truth revealed by Eq. (25) is the mechanism wherein the dispersion of the time required for acoustic energy to undergo $n$ boundary reflections (temporal variance) monotonically expands as reflections progress (as $n$ increases). Based on this property, according to the central limit theorem, when the number of reflections $n$ is sufficiently large, the distribution $P_n(t)$ asymptotically approaches a normal distribution while its variance expands.

Here, by reinterpreting the formulation of Eyring's theory through the framework of the micro model in this study, the decisive difference in mathematical structure between Eyring's theory and the present revised theory becomes evident. In Eyring's own early papers [2], as illustrated in the conceptual diagram of FIG. 10, the time distribution of each reflection order is implicitly assumed to be a rectangular distribution with a time width of $\bar{\ell}/c$. That is, the probability density function of the $n$-th order reflected sound in Eyring's theory, $P_{n,\text{Eyring}}(t)$, can be expressed as follows:

$$P_{n,\text{Eyring}}(t) = \begin{cases} c/\bar{\ell} & \text{for } n\bar{\ell}/c \le t < (n+1)\bar{\ell}/c \\ 0 & \text{otherwise} \end{cases} \tag{26}$$

FIG. 10. Conceptual diagram of the rectangular distribution implicitly assumed in Eyring's theory.

The reverberation decay model of Eyring's theory can also be perfectly expressed using the framework of the general equation (Eq. 23) in this study by employing this rectangular distribution, as shown below:

$$E_{\text{Eyring}}(t) = \frac{W}{V}\frac{\bar{\ell}}{c}\sum_{n=0}^{\infty}(1-\bar{\alpha})^n \int_t^{\infty} P_{n,\text{Eyring}}(\tau)d\tau \tag{27}$$

Comparing the proposed micro model (Eq. 23) with Eyring's micro model (Eq. 27), an immediate structural difference is the treatment of the direct sound ($n = 0$). While the proposed model explicitly isolates the direct sound due to its distinct physical property ($\bar{\ell}_d$), Eyring's formulation mathematically absorbs it into the same summation starting from $n = 0$. However, regarding the stochastic process that dominates the overall reverberation decay, it is clear that although the integration structure is identical, the core probability density function is fundamentally different. In Eyring's rectangular distribution $P_{n,\text{Eyring}}(t)$, the shape of the distribution (time width $\bar{\ell}/c$) remains constant even as the number of reflections $n$ increases, meaning the variance of the arrival time does not expand. Smoothing (making continuous) Eq. (27), which possesses this constant variance, yields nothing other than Eyring's well-known macroscopic decay formula, $E_0 \exp\left[ct \ln(1-\bar{\alpha})/\bar{\ell}\right]$.

In other words, it is revealed at the microscopic mathematical level that Eyring's theory relies on a process assuming that the time distribution of the acoustic energy at each reflection order maintains an identical distribution shape without accompanying variance expansion. This "lack of temporal variance expansion" is precisely the fundamental cause of Eyring's theory underestimating the reverberation time pointed out in the previous study [31], as will be mathematically demonstrated in the next chapter.

### C. Role of $\bar{\ell}_d$ in Scaling Temporal Variance

The standard deviation $\sigma$ of the base distribution, which determines the degree of variance expansion in the micro model, is dictated by the geometric properties of the space and the propagation characteristics of the sound waves. If the propagation time between walls for a single reflection, mentioned in Section A, is redefined here as a random variable $X$, this base distribution has a mean of $\mu = \bar{\ell}/c$, as shown in FIG. 11. As derived in Eq. (24), the direct sound fundamentally follows an exponential distribution, making its standard deviation $\sigma_d$ mathematically equal to its mean, $\bar{\ell}_d/c$. Considering this nature of spatial propagation in a diffuse sound field, it is assumed that the dispersion of the free propagation time per reflection (standard deviation $\sigma$) is also scaled using the mean residual free path $\bar{\ell}_d$ as follows:

$$\sigma = \frac{\bar{\ell}_d}{c} \qquad (28)$$

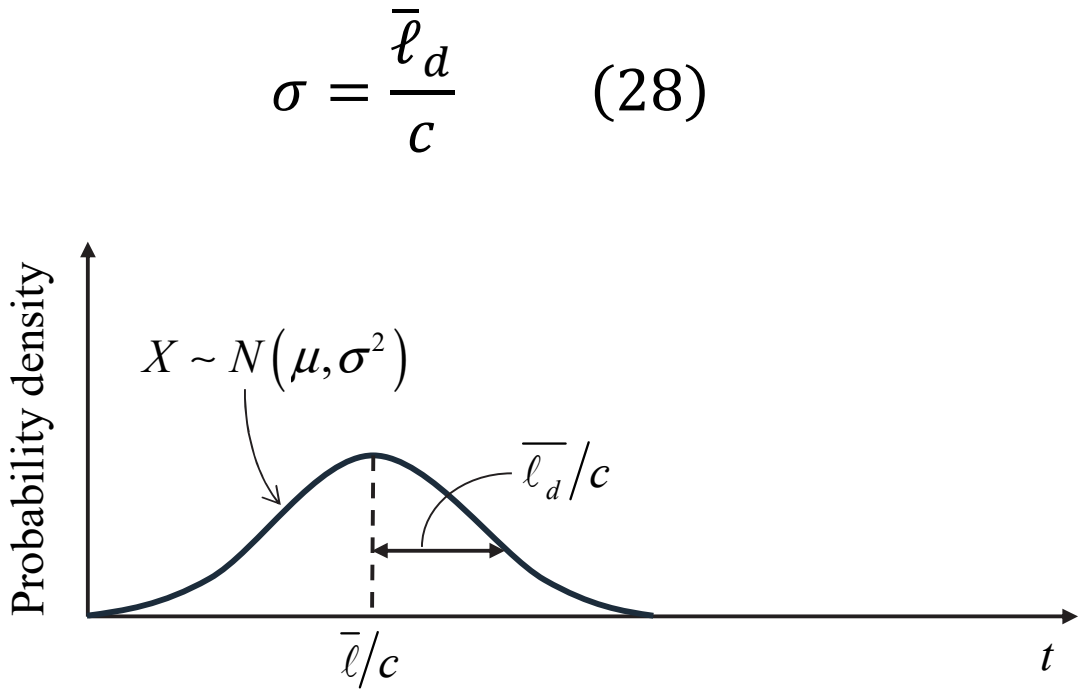

FIG. 11. Probability density function of a stochastic variable $X$ following a normal distribution $N(\mu, \sigma^2)$.

That is, the mean residual free path $\bar{\ell}_d$ is considered not only to determine the average propagation distance of the direct sound but also to serve as a "scaling factor" that determines the magnitude of the temporal variance of acoustic energy propagating uniformly throughout the space.

The validity of this mathematical assumption—that the microscopic variance $\sigma^2$ is governed by the mean residual free path $\bar{\ell}_d$—will become evident in the next chapter, where it is mathematically demonstrated that the macro model (differential equation) of Chapter IV and the micro model (sequential convolution) of this chapter share a fundamental consistency, despite being entirely different approaches.

## VI. EXPLICIT REVERBERATION FORMULA AND STRUCTURAL CONSISTENCY

This chapter provides a quantitative and mathematical evaluation of the revised reverberation theory derived from two entirely different approaches—macroscopic and microscopic—in Chapters IV and V, respectively. First, the mean residual free path of direct sound, $\bar{\ell}_d$, which was left as a variable, is specified based on physical considerations to derive a practical reverberation time formula from the general theory. The "geometric cutoff model" applied here is not merely a benchmark for theoretical comparison, but a crucial process that provides the final formula for applying the revised reverberation theory to actual acoustic design. Subsequently, a quantitative comparison is made between the derived new formula and conventional theories. Finally, it is mathematically demonstrated that the macroscopic and microscopic models share a consistent physical trend, exhibiting an exact match in the first-order approximation and settling at a similar physical scale in the second-order correction term, thereby elucidating the physical mechanism by which Eyring's theory underestimates the reverberation time.

### A. Geometric Cutoff Model and Explicit Formula

To complete the revised reverberation theory derived in this paper as a practical formula, it is necessary to specify the relationship between the mean residual free path of direct sound, $\bar{\ell}_d$, and the mean free path of the entire space, $\bar{\ell}$. In the conventional Sabine theory, it was implicitly assumed that the residual distance for a sound wave (both direct and reflected sound) propagating in a space to reach the next wall is always $\bar{\ell}$, and its time decay term was treated as $\exp(-ct/\bar{\ell})$. However, in this revised theory, this is expressed as $\exp(-ct/\bar{\ell}_d)$, strictly distinguishing the physical reality that $\bar{\ell}_d < \bar{\ell}$.

To specify $\bar{\ell}_d$ here, the maximum value of the individual direct sound path $\ell_d$ is considered from a probabilistic and statistical perspective. The expected value of this maximum distance from any point in the space to a wall, $E[\max(\ell_d)]$, can be assumed to correspond to the case where the sound source is located on a boundary wall, in which case $E[\max(\ell_d)] = \bar{\ell}$. Based on this physical constraint, this study applies a "geometric cutoff model" as the final structural component of the theory, which performs integration of Sabine's decay function $\exp(-x/\bar{\ell})$ with respect to the propagation distance $x$, using $\bar{\ell}$ as the upper limit (end point of integration). Consequently, based on the previous study [31], the relationship between the two is determined by the following equation:

$$\bar{\ell}_d = \int_0^{\bar{\ell}} \exp(-x/\bar{\ell})\, dx = \bar{\ell}(1 - e^{-1}) \qquad (29)$$

Here, $e$ is Napier's constant. Specifying $\bar{\ell}_d$ by this Eq. (29) means deriving a specific solution applicable to actual sound field prediction from the general solution, and this becomes the new reverberation time formula in this theory.

This relationship is substituted into the general formula for the time decay rate $\lambda$ in the macroscopic model derived in Chapter IV. Here, to clarify the subsequent mathematical developments, a parameter $\mu = \bar{\ell}/c = 4V/(cS)$ is defined to represent the mean time interval between successive wall reflections. The dimensionless parameter $\lambda\mu$, obtained by multiplying the time decay rate $\lambda$ by $\mu$, is expressed using the average absorption coefficient $\bar{\alpha}$ as follows:

$$\lambda\mu = \frac{\bar{\alpha}}{1 - e^{-1}\bar{\alpha}} \qquad (30)$$

By solving Eq. (30) for the time decay rate due to boundary absorption, $\lambda_{\text{boundary}}$, and adding the decay term due to air absorption introduced in Eq. (13) of Chapter IV, $\lambda_{\text{air}} = cm$ (where $m$ is the energy attenuation coefficient of air), the total time decay rate of the entire space is obtained as $\lambda = \lambda_{\text{boundary}} + \lambda_{\text{air}}$. By converting this into the reverberation time $T$ (the time required for energy to decay by 60 dB, $T \approx 13.8/\lambda$), an explicit formula representing the practical "revised reverberation time" of this theory in a diffuse sound field is derived. Expressed using the total absorption of the room (equivalent absorption area) $A = S\bar{\alpha}$, this formula takes an extremely transparent form:

$$T = \frac{55.3V}{c[A(1 - e^{-1}\bar{\alpha})^{-1} + 4mV]} \qquad (31)$$

Equation (31) is a practical form that can be directly compared with the conventional Sabine-Knudsen and Eyring-Knudsen formulas. When air absorption is neglected ($m = 0$), Eq. (31) simplifies to $T = \frac{55.3V}{cA}(1 - e^{-1}\bar{\alpha})$, revealing a concise structure in which Sabine's formula ($55.3V/cA$) is multiplied by a correction term $(1 - e^{-1}\bar{\alpha})$. In this case, it completely converges to Sabine's formula in the limit of $\bar{\alpha} \to 0$ (where the correction term becomes 1). Moreover, under the condition of complete absorption ($\bar{\alpha} = 1$), the reverberation time does not become zero, but settles at a finite value corresponding to the mean time that the direct sound remains in the space (i.e., $1 - e^{-1}$ times Sabine's formula, or approximately 63.2%).

It should be noted that the attenuation due to air absorption is an independent physical process treated additively and universally across Sabine's, Eyring's, and the present theories. Therefore, to elucidate the fundamental theoretical differences in the boundary absorption models, the air absorption term is omitted ($m = 0$) in the subsequent quantitative comparisons and mathematical verifications of consistency among the models.

### B. Quantitative Comparisons among Sabine, Eyring, and Revised Formulas

To clarify the physical and practical differences in the predicted reverberation times among the theories, a quantitative comparison of the percentage change (relative error) in reverberation time for each theory is conducted under the baseline condition where air absorption is neglected ($m = 0$), referencing both Sabine's and Eyring's theories. First, the percentage change $\Delta T/T_{\text{Sabine}}$ [%] for each theory $T_x$ relative to Sabine's reverberation time $T_{\text{Sabine}}$ is defined as follows:

$$\%\ \text{Change relative to } T_{\text{Sabine}} = \left(\frac{T_x}{T_{\text{Sabine}}} - 1\right) \times 100 \qquad (32)$$

Substituting Eyring's theory ($T_{\text{Eyring}}$) and the present revised theory ($T_{\text{Revised}}$) into $T_x$, the respective percentage changes are expressed as functions of the average absorption coefficient $\bar{\alpha}$ as follows:

$$\left(\frac{T_{\text{Eyring}}}{T_{\text{Sabine}}} - 1\right) \times 100 = \left(\frac{\bar{\alpha}}{-\ln(1-\bar{\alpha})} - 1\right) \times 100 \qquad (33)$$

$$\left(\frac{T_{\text{Revised}}}{T_{\text{Sabine}}} - 1\right) \times 100 = -100e^{-1}\bar{\alpha} \qquad (34)$$

The calculated results of Eqs. (33) and (34) are shown in FIG. 12. As is evident from FIG. 12, both Eyring's theory and the revised theory predict shorter reverberation times than Sabine's theory, with the difference widening as the average absorption coefficient $\bar{\alpha}$ increases. Notably, while the percentage change of the present revised theory exhibits a strictly linear relationship with a slope of $-100e^{-1} \approx -36.8\,\%/\bar{\alpha}$, that of Eyring's theory drops non-linearly as absorption increases.

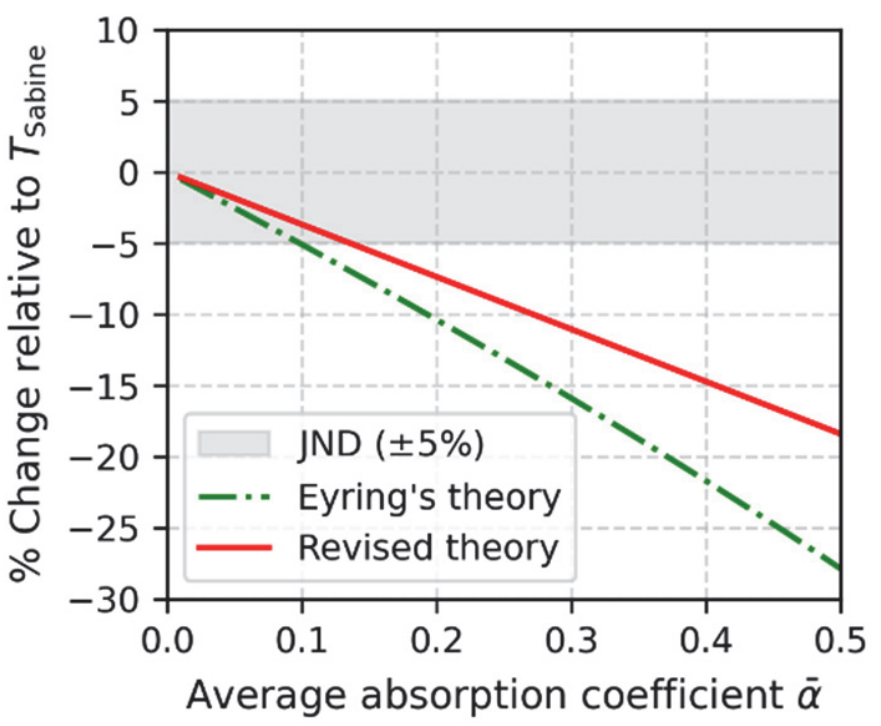

FIG. 12. Percent change in reverberation times predicted by Eyring's and the revised theories relative to $T_{\text{Sabine}}$ as a function of the average absorption coefficient $\bar{\alpha}$. The shaded region indicates the just noticeable difference (JND) threshold of $\pm 5\%$.

Similarly, the percentage change $\Delta T/T_{\text{Eyring}}$ [%] relative to Eyring's reverberation time $T_{\text{Eyring}}$, which is currently widely used in practical room acoustic design, is given by:

$$\%\ \text{Change relative to } T_{\text{Eyring}} = \left(\frac{T_x}{T_{\text{Eyring}}} - 1\right) \times 100 \qquad (35)$$

The percentage changes of Sabine's theory and the revised theory based on Eq. (35) are shown in FIG. 13. As illustrated in FIG. 13, the values predicted by Sabine's theory and the revised theory are consistently longer than those by Eyring's theory. Across the entire range of absorption coefficients, the reverberation time calculated by the present revised theory settles into an intermediate and physically consistent position: shorter than Sabine's theory and longer than Eyring's theory.

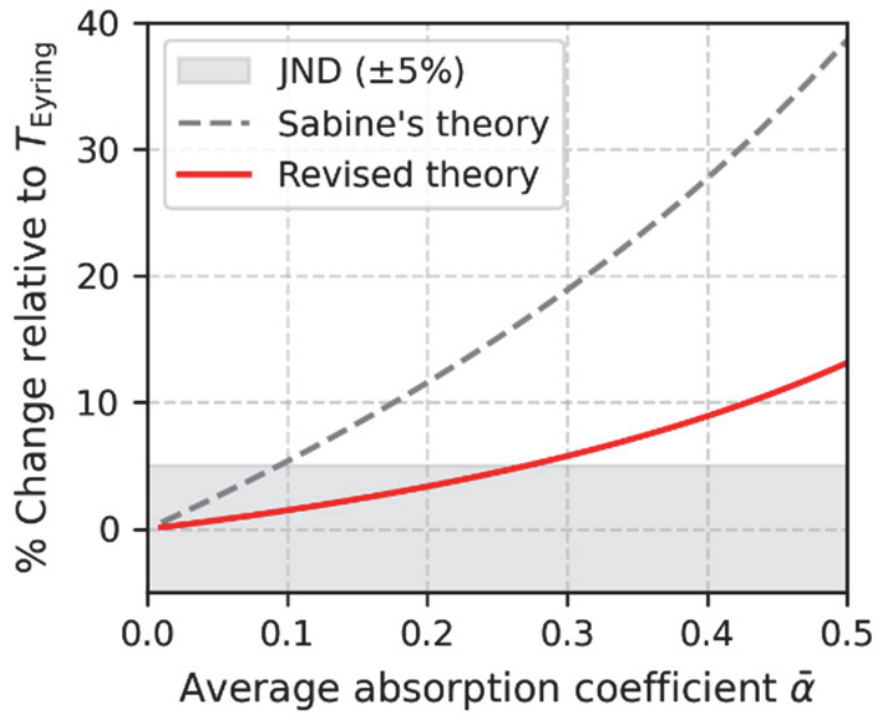

FIG. 13. Percent change in reverberation times predicted by Sabine's and the revised theories relative to $T_{\text{Eyring}}$ as a function of the average absorption coefficient $\bar{\alpha}$. The shaded region indicates the just noticeable difference (JND) threshold of $\pm 5\%$.

Next, the perceptual impact of these theoretical differences on human hearing is considered. The Just Noticeable Difference (JND) for reverberance, based on reverberation time and early decay time, is defined as approximately 5% (ranging from about 4% to 6.5%) in previous studies [39, 40] and in the international standard ISO 3382-1 [41]. Thus, a difference exceeding 5% in calculated values represents not merely a mathematical variation, but a substantial systematic error that is clearly perceptible in acoustic design.

In practical room acoustic design, typical target average absorption coefficients $\bar{\alpha}$ (in the mid-frequency octave bands, 500 Hz–1 kHz) include concert halls ($\bar{\alpha} \approx 0.18$–$0.25$), multipurpose halls, auditoriums, and classrooms ($\bar{\alpha} \approx 0.25$–$0.30$), motion picture theaters ($\bar{\alpha} \approx 0.35$–$0.45$), and studio control rooms ($\bar{\alpha} \approx 0.40$–$0.50$). The behavior of each theory across these absorption ranges is summarized as follows:

1. **JND 5% Exceedance Thresholds:** As seen in FIG. 12, the difference between Eyring's theory and Sabine's theory exceeds 5% when $\bar{\alpha} > 0.10$, whereas the present revised theory differs from Sabine's theory by 5% or more when $\bar{\alpha} > 0.14$. Furthermore, as shown in FIG. 13, the difference between the revised theory and Eyring's theory exceeds the 5% JND threshold when $\bar{\alpha} > 0.27$.

2. **Quantitative Differences in Practical High-Absorption Spaces:** In multipurpose halls and auditoriums ($\bar{\alpha} \approx 0.25$–$0.30$), the difference between the revised theory and Eyring's theory reaches the threshold region of the JND (approximately 4.5%–5.5%). In dead spaces such as motion picture theaters and control rooms ($\bar{\alpha} \approx 0.35–0.50$), the predicted values of the present revised theory are approximately 6.5%–11% longer than those of Eyring's theory, manifesting as a systematic error that clearly exceeds the JND.

The above comparison demonstrates that when Eyring's theory is applied to high-absorption spaces in practical design, it underestimates reverberation times to a degree perceptible by human hearing. The present revised theory properly corrects this systematic error arising from Eyring's unconsidered temporal variance expansion (the effect of the mean residual free path $\bar{\ell}_d$, which acts to extend the reverberation time relative to Eyring's prediction).

### C. Structural Consistency between Macroscopic and Microscopic Models

This section verifies the mathematical relationship between two entirely independent approaches: the macroscopic model based on the differential equation of acoustic energy density (Chapter IV), and the microscopic model based on the probabilistic propagation process of individual sound waves (Chapter V). If the underlying physical reality described by these two models is fundamentally identical, their mathematical behavior should be extremely close. To verify this, under the condition where air absorption is neglected ($m = 0$), the dimensionless time decay rate $\lambda\mu$ predicted by each theory is expanded as a power series (Taylor expansion) of the average absorption coefficient $\bar{\alpha}$, and the behavior of the lower-order terms is compared.

First, the dimensionless time decay rate $\lambda_{\text{macro}}\mu$ derived from the macroscopic model of this study in Eq. (30) is expanded:

$$\lambda_{\text{macro}}\mu = \frac{\bar{\alpha}}{1 - e^{-1}\bar{\alpha}} = \bar{\alpha} + e^{-1}\bar{\alpha}^2 + e^{-2}\bar{\alpha}^3 + O(\bar{\alpha}^4) \qquad (36)$$

The first-order term in Eq. (36) is $\bar{\alpha}$, which perfectly matches the decay rate of Sabine's theory in the low absorption region. The coefficient of the second-order correction term is $e^{-1} \approx 0.368$.

On the other hand, expanding the dimensionless time decay rate of the currently prevailing Eyring theory, $\lambda_{\text{Eyring}}\mu = -\ln(1 - \bar{\alpha})$, yields:

$$\lambda_{\text{Eyring}}\mu = -\ln(1 - \bar{\alpha}) = \bar{\alpha} + \frac{1}{2}\bar{\alpha}^2 + \frac{1}{3}\bar{\alpha}^3 + O(\bar{\alpha}^4) \qquad (37)$$

Comparing Eq. (36) and Eq. (37), both agree perfectly up to the first-order term. However, a clear discrepancy arises in the second-order coefficient: Eyring's theory yields $1/2 = 0.500$, whereas the proposed macroscopic model yields $e^{-1} \approx 0.368$. The fundamental reason why Eyring's theory predicts excessively short reverberation times (i.e., overestimates the decay rate) lies in this overestimation of the higher-order coefficients, particularly the second-order term.

Here, these results are compared with the analytical results of the microscopic model (convolution integral model) constructed in Chapter V. In Chapter V, the pole equation that the time decay rate $\lambda_{\text{micro}}$ must satisfy was derived from the Laplace transform of the impulse response and the analysis of its dominant pole. Let $\ln p_0(-\lambda)$ be the moment-generating function (Laplace transform) of the probability density function $p_0(t)$ for a single inter-wall propagation time $t$. With a mean $\mu = \bar{\ell}/c$ and a standard deviation $\sigma = \bar{\ell}_d/c$, the cumulant expansion around $\lambda \to 0$ can be approximated as:

$$\ln p_0(-\lambda) \approx \lambda\mu + \frac{1}{2}\lambda^2\sigma^2 \qquad (38)$$

Taking the logarithm of the microscopic pole equation $1 - (1-\bar{\alpha})p_0(-\lambda) = 0$ and substituting Eq. (38) along with the power series expansion of $-\ln(1-\bar{\alpha})$, the following equation for the dimensionless time decay rate $\lambda_{\text{micro}}\mu$ is obtained:

$$\lambda_{\text{micro}}\mu + \frac{1}{2}\left(\frac{\sigma}{\mu}\right)^2 (\lambda_{\text{micro}}\mu)^2 \approx \bar{\alpha} + \frac{1}{2}\bar{\alpha}^2 \qquad (39)$$

By expanding the dimensionless time decay rate as $\lambda_{\text{micro}}\mu = a_1\bar{\alpha} + a_2\bar{\alpha}^2 + O(\bar{\alpha}^3)$ and comparing the coefficients on both sides of Eq. (39), the first-order term in $\bar{\alpha}$ yields $a_1 = 1$, confirming a perfect match with Sabine's theory in the low absorption region. Subsequently, comparing the second-order coefficients yields $a_2 + \frac{1}{2}\left(\frac{\sigma}{\mu}\right)^2 a_1^2 = \frac{1}{2}$, from which the second-order coefficient $a_2$ is analytically derived as:

$$a_2 = \frac{1}{2}\left[1 - \left(\frac{\sigma}{\mu}\right)^2\right] = \frac{1}{2}\left[1 - \left(\frac{\bar{\ell}_d}{\bar{\ell}}\right)^2\right] \qquad (40)$$

Substituting the geometric cutoff condition of this theory, $\bar{\ell}_d = \bar{\ell}(1 - e^{-1})$, the expansion of the microscopic model becomes:

$$\lambda_{\text{micro}}\mu = \bar{\alpha} + \frac{1}{2}[1 - (1 - e^{-1})^2]\bar{\alpha}^2 + O(\bar{\alpha}^3) \approx \bar{\alpha} + 0.300\bar{\alpha}^2 + O(\bar{\alpha}^3) \qquad (41)$$

From the comparison of Eq. (36) and Eq. (41), it is confirmed that the macroscopic model (handling the energy balance of the entire space) and the microscopic model (handling the probability distribution of individual sound wave propagation distances) both perfectly agree on the first-order term $\bar{\alpha}$. Although not algebraically identical in the second-order term, they settle at extremely close physical scales of 0.368 and 0.300, respectively.

The minute difference in their second-order coefficients stems from the differing assumptions regarding the shape of the base distributions: the macroscopic model treats the mean residual free path of direct sound as a geometric cutoff in the integration over propagation distance (a product of an exponential function and a rectangular window), whereas the microscopic model models it as a normal distribution using cumulant expansion. However, the most crucial fact is that while Eyring's theory (which implicitly assumes a dispersion of $\sigma = 0$) significantly overestimates the second-order term in the expansion of the dimensionless time decay rate as $0.500\bar{\alpha}^2$, both the macroscopic and microscopic models—starting from entirely independent premises—independently demonstrate a shared structural behavior that yields similarly smaller second-order correction terms compared to Eyring's overestimation.

This analytical demonstration corroborates that the theoretical corrections proposed in this study are not mere artifacts of specific mathematical assumptions, but are the result of incorporating the long-overlooked physical phenomenon of the "reverberation of direct sound" and the new parameter of the "mean residual free path" into the models.

## VII. NUMERICAL VALIDATION BY COMPUTER SIMULATION

In this chapter, the revised reverberation theory (both macroscopic and microscopic models) deductively constructed in the previous chapters is quantitatively verified using computer simulation to determine how accurately it describes the probabilistic propagation process of sound waves in a three-dimensional geometric space.

### A. Simulation Method and Conditions

Similar to the previous study [31, 32], a Monte Carlo simulation based on the ray-tracing method was employed. In this verification, no specific sound source or receiving point was set. Instead, as an initial condition, $10^7$ acoustic particles were randomly and uniformly distributed throughout the entire room with random propagation directions to simulate an ideally diffuse sound field. The total acoustic power $W$ of the initial sound source was set to $1.0\ \mathrm{W}$, and the speed of sound $c$ was set to $340\ \mathrm{m/s}$, which enabled the calculation of the absolute energy levels and time scales of the decay curves. Starting the calculation from this state, the energy absorption and reflection at the walls (assuming perfectly diffuse reflections obeying Lambert's cosine law for all walls) were tracked, and the time variation of the total acoustic energy in the room was recorded to calculate the reverberation decay curve of the entire space. The calculation time step for tracking and recording the particles was set to $0.01/c$ seconds.

The primary objective of this verification is to purely evaluate the structural differences between Sabine's and Eyring's theories associated with wall absorption, and the corrective effect of the proposed theory. Therefore, to eliminate the masking effect of air absorption, which is additively common to all theories, the air absorption coefficient was set to $m = 0$.

Furthermore, to show that the prediction accuracy of this theory is scale-invariant (i.e., does not depend on the spatial scale $V/S$), two rectangular solid models (large and small) simulating the proportions of a shoebox-type hall were established:

- Large Model: 50 m × 20 m × 15 m
- Small Model: 20 m × 8 m × 6 m

The small model is a similar reduction of the large model by a factor of 0.4 for each side, resulting in a $V/S$ ratio that is also 0.4 times smaller. Assuming that all walls in these spaces have a uniform average absorption coefficient $\bar{\alpha}$, the decay curves were calculated by varying $\bar{\alpha}$ from 0.2 to 0.5 in increments of 0.1.

**B. Comparison of Reverberation Decay Curves**

The decay curves obtained from the simulation are compared with the theoretical decay curves according to Sabine, Eyring, and the present revised theories (macro and micro). The decay curve of the revised macroscopic theory was calculated by substituting the geometric cutoff condition $l_d$ identified in Chapter VI into the total energy decay equation derived in Chapter IV. Correspondingly, the decay curve of the revised microscopic theory was obtained by energetically summing the Schroeder integrals of the temporal distributions for both the direct sound and all reflection orders, based on the sequential convolution formulation constructed in Chapter V. The decay curves of Sabine and Eyring were calculated using their well-known fundamental formulas.

As representative conditions, the decay curves for the large and small spaces in the moderate absorption region ($\bar{\alpha} = 0.3$) and the high absorption region ($\bar{\alpha} = 0.5$) are shown in FIG. 14. As FIG. 14 illustrates, Sabine's theory (gray dashed line) decays more slowly than the simulation results (thick light gray line), overestimating the reverberation. Conversely, Eyring's theory (green dash-dot line) decays faster than the simulation, indicating a clear underestimation.

The decay curves of the revised theories proposed in this study (macro model: red solid line, micro model: blue dashed line) and the simulation almost perfectly overlap each other. This fact corroborates the mathematical analysis in Chapter VI that the two revised theories derived from entirely different approaches share a fundamental consistency, and demonstrates that the present theory captures the probabilistic propagation process in a three-dimensional geometric space with high accuracy.

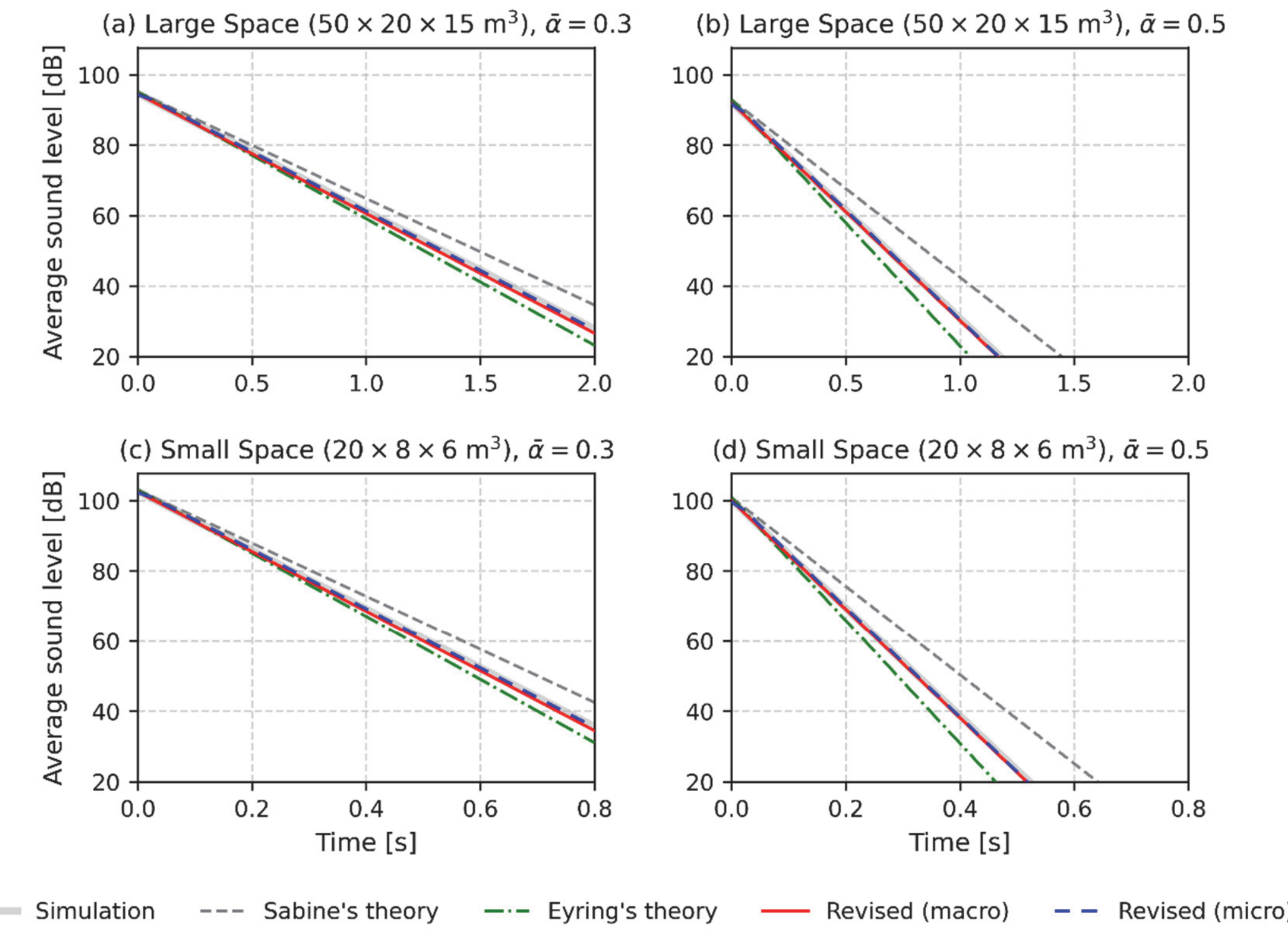

FIG. 14. Comparison of reverberation decay curves obtained from theoretical models and ray-tracing simulations for (a) Large Space (50 × 20 × 15 $m^3$) with $\bar{\alpha} = 0.3$, (b) Large Space with $\bar{\alpha} = 0.5$, (c) Small Space (20 × 8 × 6 $m^3$) with $\bar{\alpha} = 0.3$, and (d) Small Space with $\bar{\alpha} = 0.5$. The thick gray bands represent the simulation results, while the lines correspond to the theoretical predictions.

### C. Quantitative Validation of Reverberation Times and Scale-Invariant Accuracy

To further validate these visual observations, the error in reverberation time is quantified to confirm its scale-invariant accuracy. It should be noted that the simulated sound energy decay curves were obtained using Schroeder's backward integration method [36]. While the evaluation of non-exponential or multiple-slope decay curves typically requires advanced nonlinear regression approaches [42], the simulated sound fields in this study, which assume perfectly diffuse conditions, exhibited nearly ideal linear decays. Therefore, the reverberation time in the simulation could be obtained by directly extracting the time required for the energy level to decay by exactly 60 dB from the curve without ambiguity. This direct extraction method from the decay curves was equally applied to all theoretical models. Consequently, the reverberation time for the revised microscopic model, which lacks an explicit formula, was also determined directly from its theoretical decay curve. For Sabine's, Eyring's, and the revised macroscopic theories, it was confirmed that the values extracted from their respective curves perfectly matched the analytical values computed using their explicit formulas. The absolute values of the reverberation times calculated from the simulations and predicted by each theoretical model for both the large and small spaces are summarized in Table I.

TABLE I. Comparison of reverberation times $T_{60}$ (s) predicted by theoretical models and obtained from ray-tracing simulations for the Large Space ($50 \times 20 \times 15$ m$^3$) and Small Space ($20 \times 8 \times 6$ m$^3$) across different average absorption coefficients $\bar{\alpha}$.

| Space | Model / Method | $\bar{\alpha} = 0.2$ | $\bar{\alpha} = 0.3$ | $\bar{\alpha} = 0.4$ | $\bar{\alpha} = 0.5$ |
|---|---|---|---|---|---|
| Large Space ($50 \times 20 \times 15$ m$^3$) | Simulation | 2.79 | 1.79 | 1.29 | 0.98 |
| | Sabine | 2.97 | 1.98 | 1.49 | 1.19 |
| | Eyring | 2.66 | 1.67 | 1.16 | 0.86 |
| | Revised (macro) | 2.75 | 1.76 | 1.27 | 0.97 |
| | Revised (micro) | 2.80 | 1.80 | 1.29 | 0.98 |
| Small Space ($20 \times 8 \times 6$ m$^3$) | Simulation | 1.12 | 0.72 | 0.51 | 0.39 |
| | Sabine | 1.19 | 0.79 | 0.59 | 0.48 |
| | Eyring | 1.07 | 0.67 | 0.47 | 0.34 |
| | Revised (macro) | 1.10 | 0.71 | 0.51 | 0.39 |
| | Revised (micro) | 1.12 | 0.72 | 0.52 | 0.39 |

Summarizing these results in FIG. 15, the relative differences (% Error relative to Simulation) of the reverberation times obtained from each theory are shown, using the simulation results as the true baseline (0%).

In FIG. 15, despite the different spatial scales ($V/S$), the relative errors of the predictions for the large space (filled markers) and the small space (hollow markers) perfectly overlap across all evaluated theories (Sabine, Eyring, and the revised theories). This indicates that the error behavior of any of these theories is a universal property governed solely by the absorption coefficient $\bar{\alpha}$, independent of the absolute size of the space.

Furthermore, this relative difference graph visually reveals the minute discrepancy between the macro and micro models, reflecting the difference in their second-order coefficients caused by the assumptions of the base distribution shapes. Despite this slight mathematical variance, it is clearly demonstrated that the analytical predictions of both revised models show exceptional agreement with the numerical simulation results, consistently maintaining the relative difference well within $\pm 2\%$ across all tested absorption conditions.

Assuming the simulation results represent the true values, the magnitude of the error in Eyring's theory can be confirmed. Even at $\bar{\alpha} = 0.3$, Eyring's prediction is approximately 7% shorter than the simulation value, clearly exceeding the Just Noticeable Difference (JND) of 5% for human hearing. Furthermore, under the high absorption condition of $\bar{\alpha} = 0.5$, Eyring's theory underestimates the reverberation time by approximately 12–13% relative to the simulation. In other words, if Eyring's theory is used in practical acoustic design, a systematic error occurs where the actual reverberation time experienced in the constructed space becomes perceptibly longer than the calculated design value.

This critical discrepancy must be avoided, especially in room designs where reverberation must be strictly suppressed through sound absorption. The present revised theory resolves this systematic error by incorporating the "mean residual free path" of the reflected sound propagating in the space.

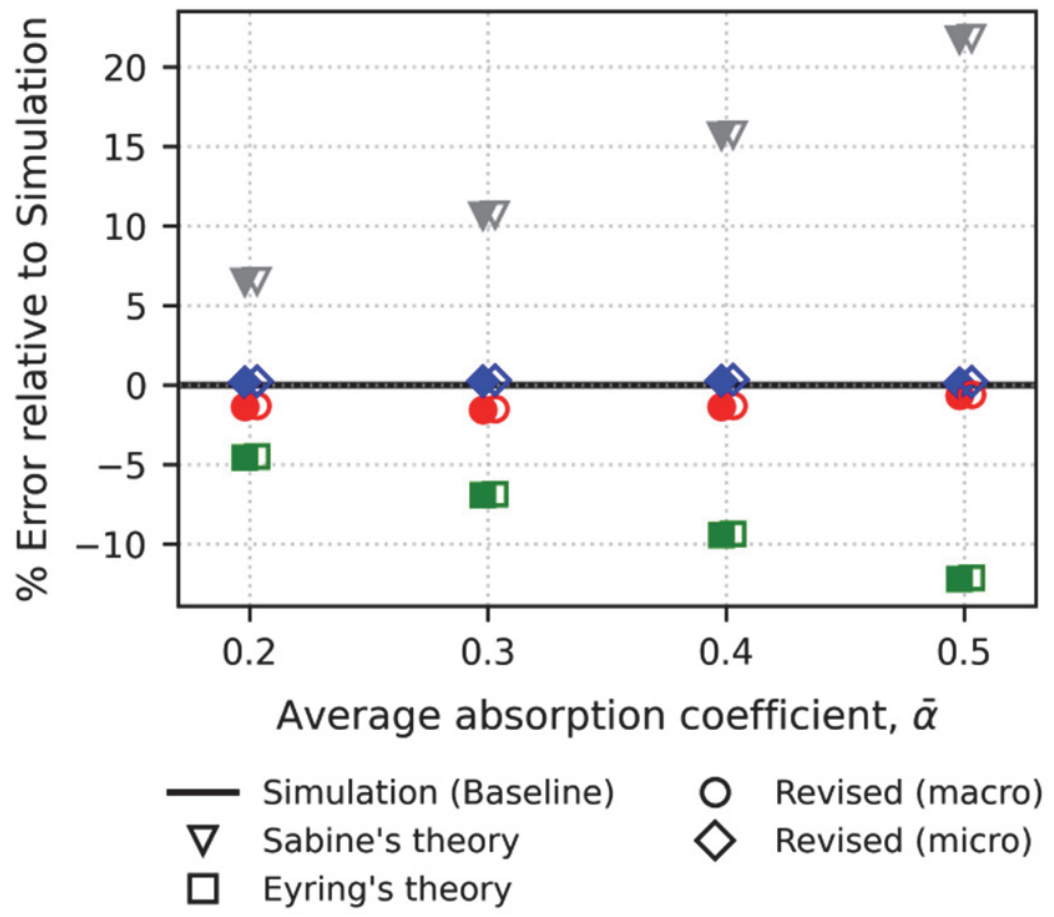

FIG. 15. Percentage errors of reverberation times predicted by theoretical models relative to ray-tracing simulations (0% baseline) as a function of average absorption coefficient $\bar{\alpha}$. Filled markers denote the Large Space ($50 \times 20 \times 15$ m$^3$), and hollow markers denote the Small Space ($20 \times 8 \times 6$ m$^3$).

### D. Physical Significance of Macroscopic Formulation and Microscopic Formulation

In FIG. 15, the reverberation time of the macro model is evaluated to be slightly shorter (less than 1%) than that of the micro model. This minute difference is a systematic error originating from the distinct mathematical structures upon which each model relies.

The macro model, based on a differential equation, assumes a continuous outflow process (Markov property) in which the acoustic energy in the space is instantaneously and perfectly diffused and mixed. In contrast, the micro model, based on sequential convolution, mathematically tracks the discrete history of sound particles flying between walls and the accompanying expansion of variance on the time axis. The reason the decay curve of the micro model perfectly matches the behavior of the Monte Carlo simulation is that this model is an analytical model that faithfully describes the discrete probabilistic propagation process in a diffuse sound field.

It is particularly noteworthy that the mean residual free path $\bar{\ell}_d = \bar{\ell}(1 - e^{-1})$, which was deductively derived from the macroscopic spatial average (geometric cutoff), completely governs the variance $\sigma = \bar{\ell}_d / c$ of the microscopic stochastic process. By modeling under the assumption that a parameter derived from macroscopic geometric constraints functions as a scaling factor for microscopic temporal variance, the results demonstrated that the model can explain the actual stochastic process of the simulation with remarkably high precision. Based on these results, it is highly reasonable to conclude that the spatial geometric constraints in a diffuse sound field and the temporal variance of acoustic energy propagation are physically almost equivalent.

## VIII. CONCLUSION

This study has established a comprehensive generalization and mathematical foundation for a revised reverberation theory, fundamentally resolving the theoretical contradictions inherent in Sabine's and Eyring's classical formulas. By introducing the core concept of the "reverberation of direct sound" and the mean residual free path $\bar{\ell}_d$, the long-standing inconsistencies in room acoustics have been structurally resolved through the following breakthroughs:

First, a generalized macroscopic model was rigorously derived directly from the fundamental energy continuity differential equation. This new formulation seamlessly integrates air absorption and the new spatial parameter $\bar{\ell}_d$, successfully separating the direct and reflected sound energies to overcome the mathematical breakdown caused by Eyring's absorption term.

Second, by reconstructing the microscopic model as a sequential convolution process of probability density functions, its exact match in the first-order approximation and qualitative agreement in the second-order term with the macroscopic model were mathematically demonstrated. This unification mathematically bridges the macroscopic continuous energy balance approach with the microscopic discrete stochastic approach, establishing a fully consistent theoretical framework.

Third, the physical and mathematical mechanism behind Eyring's underestimation of reverberation time was universally elucidated. The analysis strongly indicates that this underestimation stems directly from the structural omission of temporal variance expansion ($\sigma_n^2 = n\sigma^2$) inherent to sequential multiple reflections, a conclusion that holds true regardless of the specific probability distribution assumed.

Finally, numerical evaluations using ray-tracing simulations validated the scale-invariant accuracy of the proposed theory. In quantitative validation based on the strict definition of reverberation time, the analytical predictions of both the microscopic and macroscopic models exhibited exceptional agreement with the simulation results, consistently maintaining the relative difference well within $\pm 2\%$, demonstrating remarkable precision independent of the absolute room volume.

It is crucial to emphasize that the physical impact of this revised theory is not limited to the theoretical special case of direct sound reverberation at $\bar{\alpha} = 1$. Even under ordinary acoustic conditions where reflected sound dominates ($\bar{\alpha} < 1$), the constant intervention of the mean residual free path during propagation systematically makes the actual reverberation time longer than Eyring's prediction. As demonstrated in this study, this discrepancy is not a negligible mathematical artifact but explicitly exceeds the just noticeable difference (JND) of human hearing. Therefore, resolving the structural underestimation inherent in Eyring's theory is not merely a mathematical refinement, but a vital correction that produces an audible difference in practical room acoustic design.

In conclusion, this research updates the fundamental theory of architectural acoustics—which has long relied on empirical and macroscopic approximation formulas—into a robust, mathematically consistent framework based on microscopic stochastic processes. This generalized theory for perfectly diffuse sound fields definitively establishes the theoretical limit that future reverberation theories for non-diffuse sound fields must inherently encompass.

## ACKNOWLEDGMENTS

The author would like to express his deepest gratitude to the late Professor Sho Kimura for his lifelong academic guidance and continuous encouragement, which have profoundly supported the author's research career. Special thanks are also due to Professor Katsuaki Sekiguchi, a co-author of the foundational 1997 study that provided the initial inspiration for the microscopic stochastic approach presented in this work, for constantly teaching the author the profound depth and joy of research and exploration.

## AUTHOR DECLARATIONS

### Conflict of Interest

The author has no conflicts to disclose.

### Ethics Approval

Ethics approval is not applicable, as this study did not involve human participants or animal subjects.

## DATA AVAILABILITY

Data sharing is not applicable to this article as no new data were created or analyzed in this study.